\documentclass[conference]{IEEEtran}
\IEEEoverridecommandlockouts
\usepackage{tikz}
\usetikzlibrary{arrows.meta,positioning,fit,backgrounds,calc,shapes.geometric}
\usepackage[nocompress]{cite}
\usepackage{amsmath,amssymb,amsfonts}
\usepackage{algorithmic}
\usepackage{graphicx}
\usepackage{textcomp}
\usepackage{float}
\usepackage{xcolor}
\usepackage{url}
\def\BibTeX{{\rm B\kern-.05em{\sc i\kern-.025em b}\kern-.08em
    T\kern-.1667em\lower.7ex\hbox{E}\kern-.125emX}}
\begin{document}

\title{A Novel Compound AI Model for 6G Networks in 3D Continuum\\}

\author{

\IEEEauthorblockN{Milos Gravara}
\IEEEauthorblockA{\textit{Distributed Systems Group} \\
\textit{TU Wien}\\
m.gravara@dsg.tuwien.ac.at}

\and

\IEEEauthorblockN{Andrija Stanisic}
\IEEEauthorblockA{\textit{Distributed Systems Group} \\
\textit{TU Wien}\\
a.stanisic@dsg.tuwien.ac.at}

\and 

\IEEEauthorblockN{Stefan Nastic}
\IEEEauthorblockA{\textit{Distributed Systems Group} \\
\textit{TU Wien}\\
s.nastic@dsg.tuwien.ac.at}}

\maketitle

\begin{abstract}
The 3D continuum presents a complex environment that spans the terrestrial, aerial and space domains, with 6G networks serving as a key enabling technology. Current AI approaches for network management rely on monolithic models that fail to capture cross-domain interactions, lack adaptability, and demand prohibitive computational resources. This paper presents a formal model of Compound AI systems, introducing a novel tripartite framework that decomposes complex tasks into specialized, interoperable modules. The proposed modular architecture provides essential capabilities to address the unique challenges of 6G networks in the 3D continuum, where heterogeneous components require coordinated, yet distributed, intelligence. This approach introduces a fundamental trade-off between model and system performance, which must be carefully addressed. Furthermore, we identify key challenges faced by Compound AI systems within 6G networks operating in the 3D continuum, including cross-domain resource orchestration, adaptation to dynamic topologies, and the maintenance of consistent AI service quality across heterogeneous environments.

\end{abstract}


\section{Introduction}

The integrated 3D continuum represents a fundamental shift from traditional ground-based infrastructure, seamlessly connecting terrestrial, aerial, and space-based components into a unified ecosystem. 6G networks serve as one of the key enabling technologies for this continuum, facilitating interactions between ground networks, high-altitude platforms, unmanned aerial vehicles, and satellite constellations \cite{bahare_6g_2023}. This 3D continuum extends beyond mere connectivity to encompass distributed compute and storage resources across heterogeneous domains, creating a fabric that supports ubiquitous digital services regardless of geographical location or altitude. Through 6G's orchestration capabilities, the system promises unprecedented global coverage, enhanced reliability, and improved throughput across all segments of this multidimensional environment.

However, managing this 3D environment presents significant challenges due to the high mobility, heterogeneity, and dynamic nature of its components, particularly satellite nodes that continuously reshape network topologies in real time \cite{bhattacherjee2018gearing}. Although current AI-assisted approaches to network management predominantly rely on monolithic models, these face critical limitations within the 3D continuum context. Single-model strategies typically specialize in optimizing specific domains without effectively capturing cross-domain interactions, lack the adaptability required for rapidly changing network conditions, and often require substantial computational resources unavailable across all segments of the network. Furthermore, training and operating such large monolithic models incurs prohibitive costs in terms of computation, energy, and maintenance. System performance metrics such as latency and throughput also suffer significantly when these models attempt to handle large-scale, heterogeneous network environments, creating bottlenecks that undermine real-time decision-making capabilities.

Compound AI systems \cite{chen2025optimizingmodelselectioncompound, jain2024compositionexpertsmodularcompound, davis2024networksnetworkscomplexityclass} offer distinct advantages over these current state-of-the-art approaches. By distributing intelligence across the network, Compound AI systems enable localized decision making while maintaining global coordination, efficiently utilize heterogeneous computational resources, and adapt to the unique characteristics of each domain. This paper presents Compound AI systems to address the key management challenges of the 6G networks in the 3D continuum, demonstrating how this modular yet integrated approach can enable autonomous adaptation while maintaining consistent service quality and operational efficiency across terrestrial, aerial, and space domains.


\title{\textbf{Compound AI Systems}}
\date{}
\maketitle

\section{Overview of Compound AI Systems}

\subsection{Towards system design approach to AI}


Rather than relying on a single model to handle all aspects of a complex problem, Compound AI systems decompose tasks into manageable sub-components. This modular approach enables developers to leverage specialized models or tools for different sub-tasks and control information flow. For example, lightweight models can make real-time decisions at terrestrial base stations, while different specialized models simultaneously optimize aerial and satellite resource allocation. Supporting infrastructure components such as vector databases for similarity search, orchestration frameworks for module coordination, API gateways for external tool access, and monitoring systems for quality assurance create a rich ecosystem of components to build upon. This diversity allows the system to function effectively even when parts have varying computational capacities or connectivity constraints. Similarly, routing inputs to different-sized models based on task complexity and overall system load can optimize computational resources while maintaining quality of service across heterogeneous network segments. As network conditions evolve, individual AI components can be updated independently, preserving adaptability without requiring retraining of the entire system.

\subsection{Example Compound AI System and Main Design Principles}

To better understand the concept of Compound AI systems, we turn to the following explanatory example. 

\begin{figure}[H]
    \centering
    \includegraphics[width=\columnwidth]{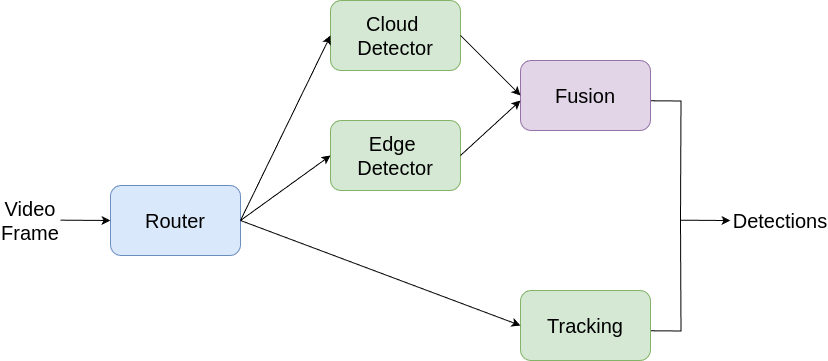}
    \caption{VATE: A Compound AI System for Edge-Cloud Object Detection and Tracking. \cite{10707231}}
    \label{fig:VATE}
\end{figure}

Fig \ref{fig:VATE} illustrates a Compound AI system for edge-cloud collaborative object detection and tracking. It demonstrates how multiple specialized modules collaborate in a unified system. At the edge, a lightweight detector identifies potential objects, while a tracking module maintains object persistence across frames. On the cloud side, a more powerful detector offers greater accuracy for challenging cases. A fusion module combines detections from both edge and cloud sources to create an improved understanding of the scene. Furthermore, a dedicated orchestrator module makes intelligent decisions about when to process input locally, when to offload to the cloud, and when to rely on tracking rather than detection, balancing between computational efficiency and accuracy.


With these considerations, we define Compound AI systems as systems composed of specialized, interoperable modules that collectively address complex AI tasks. Each module performs a distinct function and interacts via well-defined defined interfaces. The following key characteristics define the structural and functional principles that underpin Compound AI systems, enabling them to address complex tasks in a scalable and efficient manner:

\begin{itemize}
  \item \textbf{Modularity} – Separate parts of Compound AI systems can be developed, tested and maintained independently while minimizing impacts on the overall system. This represents a separation of concerns that facilitates parallel development and iterative improvement by specialized teams \cite{zhang2025flowstatehumansenabling}.

  \item \textbf{Adaptability} – The modular design of Compound AI systems enables rapid adaptation to new requirements or changing conditions by allowing individual components to be replaced, enhanced, or reconfigured without rebuilding the entire system.

  \item \textbf{Abstraction} – Internal complexities of modules are hidden behind well-defined interfaces, ensuring that changes to a module's implementation don't affect other components. This creates a clear separation between what a module does and how it accomplishes its task.

  \item \textbf{Interaction-defined Architecture} – The architecture of a Compound AI system is fundamentally defined by how modules interact with each other. These interactions establish the data flow through the system and determine how information is processed, transformed and utilized across modules.

  \item \textbf{Cost-Effectiveness} - Designing with cost-effectiveness in mind ensures that resource utilization, energy consumption, and computational requirements are optimized. This principle is crucial not only for reducing operational expenses but also for enabling an architecture where individual components can be updated or replaced without necessitating a complete retraining of the entire system
\end{itemize}

\section{Compound AI System Model}

\subsection{Definition and Core Components}

\begin{figure}[H]
  \centering
  \includegraphics[width=\columnwidth]{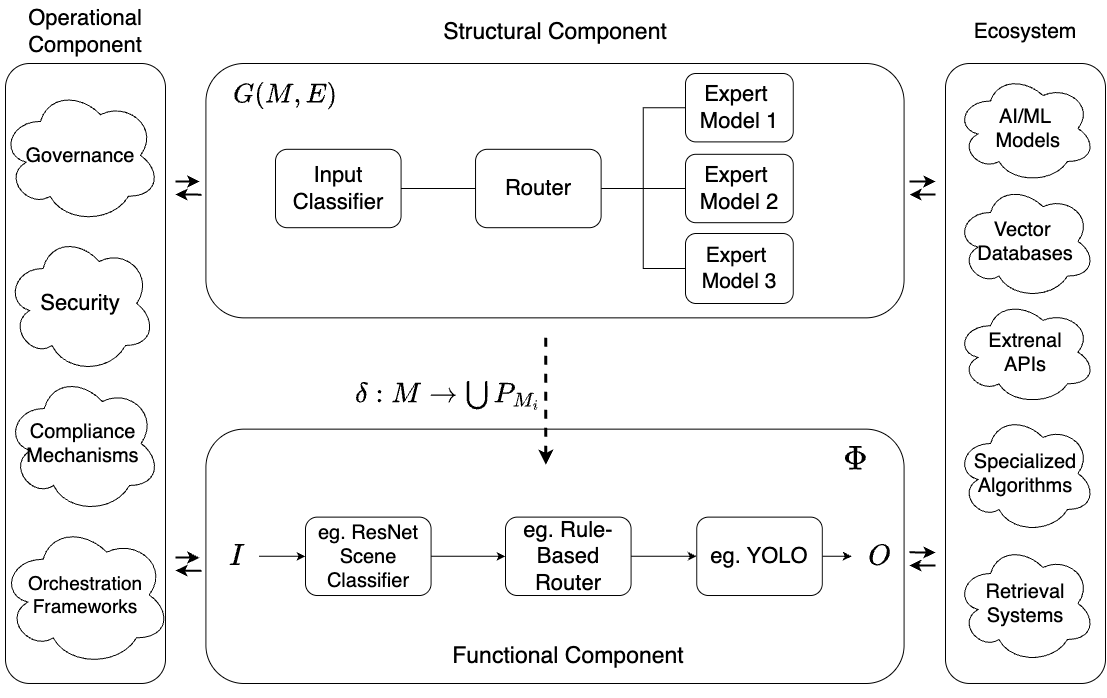}
  \caption{Tripartite model for Compound AI systems}
  \label{fig:model}
\end{figure}

In Fig \ref{fig:model} we present a model and reference architecture for Compound AI system. This reference architecture offers a conceptual framework through which to analyze, conceptualize, and implement Compound AI (CAI) systems as multi-modular, interactive, and adaptive entities. These systems are composed from a comprehensive ecosystem encompassing various AI/ML models, vector databases, retrieval mechanisms, external APIs, and domain-specific algorithmic solutions.

\vspace{0.1cm}

We define a Compound AI System formally as a triple:

\[
\text{CAI} = (S, F, O)
\]

Where:
\begin{itemize}
  \item $S$ represents the \textbf{Structural} component
  \item $F$ represents the \textbf{Functional} component
  \item $O$ represents the \textbf{Operational} component
\end{itemize}

While each component serves a distinct role, they interact in important ways:

\begin{enumerate}
  \item The Structural component ($S$) defines what modules exist and how they connect, constraining the possible implementations in the Functional component ($F$).
  \item The Functional component ($F$) realizes the abstract architecture defined in $S$ through concrete implementations, potentially informing structural changes based on implementation constraints.
  \item The Operational component ($O$) monitors and maintains $S$ and $F$, providing feedback for optimization and adaptation.
\end{enumerate}

The tripartite model naturally supports key system design principles such as modularity, adaptability, and abstraction. By clearly separating architectural, behavioral, and operational concerns, it enables more systematic approaches to the development, deployment, and maintenance of complex, multi-model AI systems.

\subsection{Structural Component (S)}

The Structural component defines the high-level architecture of the system, specifying what each module does and how information flows between modules. Formally, we represent this as a directed graph:

\[
S = G(M, E)
\]

Where:
\begin{itemize}
  \item $M = \{M_1, M_2, \ldots, M_n\}$ is the set of modules
  \item $E \subseteq M \times M$ is the set of connections between modules
\end{itemize}

For any two modules $M_i$ and $M_j$, an edge $E(M_i, M_j)$ indicates that the output of module $M_i$ serves as an input to module $M_j$.

Each module $M_i$ is characterized by its input space $I_i$ and output space $O_i$, which define the types and formats of data the module can accept and produce. The Structural component thus establishes a ``schema'' for the Compound AI system, defining the roles and relationships of its constituent parts without specifying their implementation details.

\subsection{Functional Component (F)}

The Functional component defines how the system behaves by mapping the abstract architecture to concrete implementations. It encompasses:

\[
F = (P, \Phi, \delta)
\]

Where:
\begin{itemize}
  \item $P = \{P_1, P_2, \ldots, P_n\}$ is a implementation set where each $P_i$ is an implementation pool for module $M_i$
  \item $\Phi$ is the composition function that determines the overall system behavior
  \item $\delta$ is the mapping function that selects specific implementations from pools
\end{itemize}

For each module $M_i$, the implementation pool $P_i = \{f_{i1}, f_{i2}, \ldots, f_{ik}\}$ contains multiple possible implementations that fulfill the same functional role but may differ in their performance characteristics, resource requirements, or other properties.

The mapping function $\delta: M \rightarrow \bigcup P_i$ selects a specific implementation for each module, such that $\delta(M_i) \in P_i$.

The composition function $\Phi$ integrates these implementations according to the structural blueprint to produce the overall system behavior:

\[
\Phi: I \rightarrow O
\]

Where $I$ is the input space and $O$ is the output space of the entire system.

Critically, this function ensures that the compound system's behavior matches what would be expected from a monolithic model:

\[
y = \Phi(x) \quad \text{where } x \in I,\, y \in O
\]

The composition function can be formally defined in terms of the graph execution:

\[
\Phi(x) = \Psi(G(M, E), \delta(M_1), \delta(M_2), \ldots, \delta(M_n), x)
\]

Where $\Psi$ is a graph execution function that propagates inputs through the implementation graph according to the connection pattern defined in $S$.

\subsection{Operational Component (O)}

The Operational component encompasses the infrastructure and processes that enable, maintain, and optimize the running system. This includes:

\[
O = (Mon, Sec, Gov, Orch)
\]

Where:
\begin{itemize}
  \item $Mon$ represents monitoring and observability systems
  \item $Sec$ represents security and compliance mechanisms
  \item $Gov$ represents governance frameworks and policies
  \item $Orch$ represents orchestration and resource management
\end{itemize}

The Operational component serves as the foundation that supports both the Structural and Functional components, providing:

\begin{enumerate}
  \item \textbf{System monitoring} that tracks performance, resource utilization, and failure modes
  \item \textbf{Security controls} that protect system integrity and data privacy
  \item \textbf{Governance mechanisms} that ensure compliance with regulations and ethical standards
  \item \textbf{Orchestration tools} that manage deployment, scaling, and resource allocation
\end{enumerate}

This component parallels DevOps and MLOps practices in software engineering but extends them to address the unique challenges of compound AI systems.

\section{Open Challenges for Compound AI in 6G Networks for 3D Continuum}

To realize our vision of Compound AI, the following challenges need to be addressed.

\subsection{Cross-Domain Resource Orchestration}
Orchestrating Compound AI resources across the 3D continuum faces unique constraints that existing AI solutions fail to address. Terrestrial components can leverage high computational capacity but are limited in coverage, aerial platform-based components faces energy and computational constraints, while satellite-hosted systems can provide wide coverage but with significant computational and latency limitations. Current AI orchestration approaches treat these domains separately, creating inefficiencies at domain boundaries. Compound AI for 6G networks requires intelligent decomposition and coordination mechanisms that can distribute AI tasks optimally across these diverse domains while accounting for their unique characteristics and computational limitations.

\subsection{Adaptation to Dynamic Network Topologies}
The constantly evolving network topologies of the 3D continuum challenge traditional deployment strategies. As satellite constellations orbit, aerial platforms move, and terrestrial demand shifts, Compound AI systems must continuously reconfigure themselves to maintain performance. Current AI composition algorithms struggle with this dynamism, leading to suboptimal configurations where component distribution becomes misaligned with actual network conditions. Compound AI for 6G networks requires adaptive systems capable of predicting topology changes and proactively reconfiguring  component distribution and communication patterns, thus maintaining overall model performance while optimizing system performance metrics.

\subsection{Maintaining AI Service Consistency}
Delivering consistent AI service quality across the 3D continuum presents significant technical challenges. As AI requests and data transition between terrestrial, aerial, and space segments, maintaining continuity of AI inference quality becomes increasingly difficult. Current approaches typically react to AI service degradations after they occur, particularly at domain boundaries where computational resources vary dramatically. Compound AI for 6G networks requires predictive capabilities that can anticipate performance variations across the continuum and implement proactive measures in order to maintain service level objectives despite the inherent heterogeneity and dynamic nature of the underlying 3D infrastructure.

\subsection{Balancing the Trade-offs}
Compound AI systems present trade-offs between improving model performance and maintaining system efficiency \cite{chaudhry2025resourceefficientcompoundaisystems}. In scenarios requiring advanced reasoning, adding specialized components to a monolithic model can enhance accuracy, precision, and recall. However, this often leads to increased latency, energy consumption, and computational demands. Conversely, when deploying on the edge, Compound AI aims to retain comparable model performance while significantly improving system efficiency to suit resource-constrained environments. These trade-offs become especially challenging across the 3D continuum, which spans from powerful data centers to limited-capacity edge devices. Addressing this complexity requires adaptive frameworks capable of dynamically reconfiguring Compound AI systems based on fluctuating network conditions, available resources, and application-specific requirements. Such adaptability would allow systems to strike the right balance between model and system performance, depending on the deployment context. Advancing this research is key to ensuring that the benefits of Compound AI outweigh the added complexity it introduces in real-world scenarios.

\section{Conclusion}

In this paper, we introduced the concept of Compound AI systems and presented a formal tripartite model that captures their structural, functional, and operational dimensions. By breaking down complex tasks into specialized, interoperable modules, Compound AI systems offer a scalable and adaptable alternative to traditional monolithic AI architectures. We explored how this general framework can address the unique challenges of 6G networks in the 3D continuum, highlighting the advantages of modularity and distributed intelligence in such complex scenarios.

Looking ahead, our future work will focus on applying this general Compound AI system model to real-world 6G use cases and other dynamic, distributed environments. This includes implementing adaptive orchestration strategies, developing proactive reconfiguration mechanisms, and validating system performance across varying network and resource conditions. By doing so, we aim to demonstrate how Compound AI systems can be effectively deployed to meet the demands of next-generation networks and beyond.

\vspace{0.2cm}

\textbf{Acknowledgments}: This research was partially funded by the EU's Horizon Europe Research and Innovation Program as part of the NexaSphere project (GA No. 101192912).

\bibliographystyle{plain} 
\bibliography{references} 

\end{document}